\documentstyle[twoside,fleqn,espcrc2]{article}

\newcommand{\AmS}{{\protect\the\textfont2
  A\kern-.1667em\lower.5ex\hbox{M}\kern-.125emS}}

\title{Black Hole Entropy in Induced Gravity
       and Information Loss}

\author{D.V. Fursaev\address{Bogoliubov Laboratory of Theoretical 
        Physics, 
        Joint Institute for Nuclear Research, \\ 
        141 980 Dubna, Russia}}
       
\begin{document}

\begin{abstract}
The basic assumption of the induced gravity approach is
that Einstein theory is an effective, low energy-form of 
a quantum theory of constituents. 
In this approach the Bekenstein-Hawking
entropy $S^{BH}$ of a black hole can be interpreted 
as a measure of the 
loss of information about constituents inside the black hole horizon.  
To be more exact, $S^{BH}$ is determined by quantum correlations 
between "observable" and "non-observable" states with positive and
negative energy $\cal E$, respectively.  It is important that for 
non-minimally coupled constituents $\cal E$  differs from the canonical 
Hamiltonian $\cal H$.  This explains why previous definitions of the 
entanglement entropy in terms of $\cal H$ failed to reproduce $S^{BH}$.  
\end{abstract}

\maketitle

The idea of induced gravity has been formulated 
first by A.D.Sakharov \cite{Sakharov}.
Its basic assumption is that
the dynamical equations for the gravitational field $g_{\mu\nu}$ 
are related to properties of the physical vacuum
which has a microscopic structure like a usual
medium. A suitable example for the comparison is a crystal lattice.  
The metric $g_{\mu\nu}$ plays a role similar to a macroscopic 
displacement in the 
crystal and gravitons are collective excitations of 
the microscopic constituents of the vacuum analogous to phonons.  In 
the crystal, a deformation results in the change of the energy 
expressed in terms of displacements.  In a similar way, the physical 
vacuum responses to variations of $g_{\mu\nu}$ by changing its energy 
due to polarization effects.  These quantum effects can be described by 
the effective action $\Gamma[g_{\mu\nu}]$ for the metric
\begin{equation}\label{1}
e^{i\Gamma[g_{\mu\nu}]}=
\int [D\Phi]e^{iI[\Phi,g_{\mu\nu}]}~~~.
\end{equation}
The notation $\Phi$ stands for the constituent fields, and
$I[\Phi,g_{\mu\nu}]$ denotes their classical action.
Sakharov's idea is based on the observation 
that one of leading contributions to $\Gamma[g_{\mu\nu}]$
has a form of the Einstein action. 

Why the idea of induced gravity has something to do with the
problem of the black hole entropy? The Bekenstein-Hawking entropy
\begin{equation}\label{2}
S^{BH}={1 \over 4G}{\cal A}~~~,
\end{equation}
where $\cal A$ is the area of the horizon and $G$ is
the Newton constant, appears under analysis of the laws of black hole
mechanics in classical Einstein theory. The induced gravity
is endowed with a microscopic structure, constituents,
which can be related naturally to degrees of freedom responsible for
$S^{BH}$. As was first pointed out in \cite{Jacobson},
the mechanism which makes explanation of $S^{BH}$ in
induced gravity possible is the quantum entanglement
\cite{BKLS}.
The constituents in the black hole exterior are in the mixed state
due to the presence of the horizon.
This state is characterized by the entropy
$S$ of entanglement.
In a local field theory correlations between external
and internal regions are restricted by the horizon surface, and
so $S={\cal A}/l^2$. The constant 
$l$ is the same ultraviolet cutoff which is required
to make effective action (\ref{1}) finite \cite{SU}.
Because the induced Newton constant $G$
in $\Gamma[g_{\mu\nu}]$ is proportional to $l^2$
entropy $S$ is comparable to $S^{BH}$. 

To understand whether $S^{BH}$ is a
"purebred" entanglement entropy  one has to 
avoid the cutoff procedure and
consider a theory which is free from the divergences.
The Bekenstein-Hawking
entropy is a pure low-energy quantity.
It was suggested in \cite{FFZ} that the mechanism of generation
of $S^{BH}$ is universal  
for the theories where gravity is
completely induced by quantum effects, as in Sakharov's approach
or in string theory,
and has ultraviolet finite induced Newton
constant $G$. 
The last requirement makes the theory self-consistent and
enables one to carry out the computations.
The universality means that $S^{BH}$ may be understood 
without knowing details of quantum gravity theory \cite{FFZ}
(see also \cite{Carlip} for a different realization of this 
idea).

The models which obey these requirements were constructed in
\cite{FFZ,FF:97,FF:98b,FF:99a,FF:99b}.
The constituents are chosen there as non-interacting scalar, spinor and 
vector fields. They have different masses comparable to
the Planck mass $m_{Pl}$. The masses and other parameters 
obey certain constraints which
provide cancellation of the leading (or all \cite{FF:99b}) 
divergences in effective gravitational action (\ref{1}).
In the low-energy 
limit when curvature $R$ of the background is small
as compared to $m^2_{Pl}$
\begin{equation}\label{3}
\Gamma[g_{\mu\nu}]\simeq {1 \over 16\pi G} \int \sqrt{-g}d^Dx
(R-2\Lambda)~~~.
\end{equation} 
The induced Newton constant $G$ and the cosmological
constant $\Lambda$ are the functions of parameters of the
constituents and can be computed. In four dimensions one 
imposes further restrictions to eliminate the cosmological term.
To study charged black holes one has to generalize Sakharov's
proposal and consider the induced Einstein-Maxwell theory 
(see the details in \cite{FF:99b}).

All above models include
bosonic constituents (scalars or vectors)
with a positive non-minimal coupling to the background curvature.
The presence of such couplings is unavoidable if one
wants to eliminate the divergences in $G$.
The Bekenstein-Hawking entropy of a non-extremal 
stationary black hole
has the following form
\cite{FFZ,FF:97,FF:98b,FF:99a,FF:99b}
\begin{equation}\label{4} 
S^{BH}={1 \over 4G}{\cal A}=S-{\cal Q}~~~,
\end{equation}
\begin{equation}\label{5}
S=-\mbox{Tr}\rho \ln \rho~~~,
~~~\rho={\cal N}e^{-\beta_H {\cal H}}~~~.
\end{equation}
Here $S$ is the thermal entropy of the constituents 
propagating in Planck region near the black hole horizon and
$\beta_H^{-1}={\kappa \over 2\pi}$ is the Hawking temperature. 
The operator ${\cal H}$
is the canonical Hamiltonian which generates translations
along the Killing vector field $\xi$ ($\xi^2=0$ on the horizon).  
Quantity $\cal Q$ is the vacuum average of an 
operator which has the form  of the integral over the 
horizon $\Sigma$.
$\cal Q$ is determined only by the non-minimal couplings of 
the constituents and it can be interpreted as a Noether charge 
on $\Sigma$
defined with respect to field $\xi$ \cite{FF:97,F:98b}.  
Formula (\ref{4}) is absolutely universal in  a sense
it depends neither on the choice of constituent species
nor on a type of a black hole. The entropy $S$ is always positive
and divergent and so does $\cal Q$ in the considered models.
Subtracting $\cal Q$ in (\ref{4}) compensates 
divergences of $S$ rendering $S^{BH}$ finite. 
Thus, non-minimal couplings provide both
finiteness of the effective action and of $S^{BH}$.

The physical meaning of $\cal Q$ becomes evident
if we analyze the classical energy
of the non-minimally coupled constituents defined
on a space-like hypersurface $\Sigma_t$ which crosses
the bifurcation surface $\Sigma$
\begin{equation}\label{6}
{\cal E}=\int_{\Sigma_t}T_{\mu\nu}\xi^\mu d\Sigma^\nu~~~.
\end{equation}
Here $T_{\mu\nu}$ is the stress-energy tensor and
$d\Sigma^\mu$ is the future-directed vector 
of the volume element on $\Sigma_t$. 
For the fields in black hole 
exterior \cite{F:98b} 
\begin{equation}\label{7} 
{\cal E}={\cal H}-{\kappa \over 2\pi}{\cal Q}~~~,
\end{equation} 
where $\cal H$ is the canonical Hamiltonian.
The difference between the two energies is a total divergence which 
results in boundary terms.
$\cal Q$ is the contribution 
from the horizon $\Sigma$ which is an internal
boundary of $\Sigma_t$.   
The energies $\cal H$ and $\cal E$
play different roles: $\cal H$ is the generator
of canonical transformations along $\xi$, while $\cal E$ is 
the observable energy which is related to thermodynamical 
properties of a black hole. Indeed, there is the
classical variational formula \cite{F:98b}
\begin{equation}\label{8} 
\delta M={\kappa \over 2\pi} \delta S^{BH}+\delta {\cal E}~~
\end{equation} 
which relates the change of the black hole mass at
infinity $\delta M$ to the change of the entropy 
$\delta S^{BH}$ under a
small perturbation of the energy $\delta{\cal E}$ of matter fields.

Suppose now that the total mass $M$ is fixed. Then,
the change of $S^{BH}$ is completely  determined by
change of the energy $\cal E$
in the black hole exterior. 
In classical processes the matter is
falling down in the black hole. It results in decreasing of the energy
$\delta {\cal E}<0$  and increasing of the black hole entropy $S^{BH}$.
In quantum theory $S^{BH}$ may decrease due to the
Hawking effect. In such a process $\delta {\cal E}>0$,
and a black hole absorbs a negative energy.

We can now make two observations.
First, according to (\ref{5}),  changes of $S$ 
are related to changes of the canonical energy $\cal H$,
and because $\cal H$ and $\cal E$ are different, the two
entropies, $S^{BH}$ and $S$, should be different as well.
Second, the conclusion whether a physical process 
results in a loss of the information inside the black hole
can be made by analyzing the sign of the energy change 
$\delta {\cal E}$. 
Note that using for this purpose
the sign of the canonical energy  
would result in a mistake. For instance,
if ${\cal Q}>0$, there may be 
classical processes where ${\cal H}$ and $S^{BH}$ increase
simultaneously, see (\ref{7}), (\ref{8}).

It is known that the thermal entropy $S$, defined in (\ref{5}), 
can be interpreted as an entanglement entropy \cite{Israel}.
This is true when one is interested in correlations
(entanglement) of states characterized by the certain
canonical energy $\cal H$. 
However, to describe a real information loss one has to
study correlations between
observable and non-observable states which have
positive and negative energy $\cal E$, respectively.

Consider a pair creation process
near the black hole horizon. 
The antiparticle of a pair 
carries a negative energy. It tunnels through 
the horizon and becomes non-observable. 
As a consequence, the observed particle 
is in a mixed state. The entropy which describes the corresponding
information loss can be defined by the Shannon formula
\begin{equation}\label{9}
S^I=-\sum_{\cal E}p({\cal E})\ln p({\cal E})~~~.
\end{equation}
Here $p({\cal E})$ is the probability to observe a particle with 
the certain energy $\cal E$. Obviously, $p({\cal E})$ coincides with 
the probability for the antiparticle with energy $-\cal E$
to tunnel through the horizon. 
This process can be also described in another form. 
Consider a Kruskal extension of 
the black hole space-time near the horizon. 
Then the surface $\Sigma_t$ 
consists of the two parts, $\Sigma_t^R$ (right) and $\Sigma_t^L$ 
(left), which are separated by the bifurcation surface $\Sigma$.  The 
creation of a pair is equivalent to quantum tunneling 
of a particle moving backward in time from the left region
to the right.
The corresponding amplitude is the 
evolution operator from $\Sigma_t^L$ to $\Sigma_t^R$ in Euclidean
time $\tau$. In the Hartle-Hawking vacuum $\tau=\beta_H/2$.

How to describe the tunneling probability?  
First, note that the energy has a form of the Hamiltonian
modified by a "perturbation" $\cal Q$.
Such a modified Hamiltonian may be considered
as an operator which
generates evolution of states having certain energies.
Then, the operator 
\begin{equation}\label{10}
T=e^{-{\beta_H \over 2}{\cal E}}~~~
\end{equation} 
corresponds to the tunneling transition.
Second, because a particle on $\Sigma_t^L$
is moving backward in time, 
$\cal E$ should be defined by (\ref{6})
with the past-oriented Killing field $\xi$.
It means that relation (\ref{7}) 
should be written now as
\begin{equation}\label{6l} 
{\cal E}={\cal H}+{\kappa \over 2\pi}{\cal Q}~~~,
\end{equation} 
if one leaves for $\cal Q$ the former expression.
(So $\cal Q$ is positive for positive non-minimal
couplings.)
The same relation is true for the energy and Hamiltonian
of an antiparticle
tunneling from $\Sigma_R^t$ to $\Sigma_L^t$.

The probability $p({\cal E})$ which involves transitions from all states
in one region to a state 
with the energy $\cal E$ in the other region is
\begin{equation}\label{11}
p({\cal E})
={\cal N}\langle {\cal E}| 
e^{-(\beta_H{\cal H}+{\cal Q})}|{\cal E}\rangle~. 
\end{equation}
Here $\cal Q$ is the operator which corresponds to the 
Noether charge. 
We now have to calculate $S^I$ by using (\ref{9}) and (\ref{11}).
To this aim it is convenient to introduce a free energy 
$F^I(\beta)$ 
\begin{equation}\label{12}
e^{-\beta F^I(\beta)}=\mbox{Tr}e^{-\beta {\cal E}} ~~~,
\end{equation}
which makes it possible to rewrite $S^I$ as
\begin{equation}\label{13}
S^I=\left.\beta^2 \partial_\beta F^I(\beta)\right|_{\beta=\beta_H}~~~.
\end{equation}
Our last step is to calculate $F^I(\beta)$. This can be done
by using our interpretation of $\cal Q$ as a perturbation. 
Then in the linear order
\begin{equation}\label{14}
F^I(\beta)=F(\beta)+\beta_H^{-1}\langle {\cal Q} \rangle_\beta~~~,
\end{equation}
\begin{equation}\label{15}
e^{-\beta F(\beta)}=\mbox{Tr}e^{-\beta {\cal H}} ~~~,
\end{equation}
\begin{equation}\label{16}
\langle {\cal Q} \rangle_\beta=
\mbox{Tr}({\cal Q}e^{-\beta {\cal H}})
e^{\beta F(\beta)}
~~~.
\end{equation}
Here $F(\beta)$ is the standard free energy
for a canonical ensemble at the temperature $\beta^{-1}$.
$\langle \cal Q \rangle_\beta$ is the 
thermal average of the Noether charge in this canonical ensemble.
As has been argued in \cite{FF:97}, this average is determined
by contributions of particles with negligibly small
canonical energies $\omega$ (the soft modes). The density $n_\omega$
of Bose particles at the temperature $\beta^{-1}$
is singular and proportional to the
temperature, 
$n_\omega\simeq (\beta \omega)^{-1}$.
Thus,
\begin{equation}\label{17}
\langle {\cal Q} \rangle_\beta=
{\beta_H \over \beta}\langle {\cal Q} \rangle_{\beta_H}
~~~,
\end{equation}
where
$\langle {\cal Q} \rangle_{\beta_H}\equiv {\cal Q}$ is the average in 
the Hartle-Hawking vacuum. By using (\ref{13}), (\ref{14}) and 
(\ref{15}) one finds 
\begin{equation}\label{18} 
S^I=\left.\beta^2 \partial_\beta F(\beta)\right|_{\beta=\beta_H} 
-\langle {\cal Q} \rangle_{\beta_H}= S-Q~~~.  
\end{equation}
Here we used the fact that $F(\beta)$ determines thermal entropy
(\ref{5}). Equation (\ref{18})  proves 
our interpretation of the Bekenstein-Hawking
entropy in induced gravity (\ref{4}) as the
entropy related to the information loss.

We conclude with the several remarks.

1) In induced gravity $S^{BH}$
counts only the states of the constituents
located near the horizon, the "surface 
states". Thus, 
as was argued in \cite{Sorkin},
$S^{BH}$ should not be related to 
internal 
states of the black hole which cannot influence outside.

2) Our arguments are applied to the entropy of Rindler observers
and give for this quantity the finite value $(4G)^{-1}$ per
unit surface.

3) It was suggested in \cite{FF:97} to relate
$S^{BH}$ to degeneracy
of the spectrum of the mass of the black hole $M_H$ measured
at the horizon.
If mass at infinity $M$ is fixed, change of $M_H$
coincides with the energy of an antiparticle
absorbed by a black hole in the course of the pair creation process. 
According to arguments above, this energy has to be determined by 
(\ref{6l}) and it is the operator that appears in density matrix 
(\ref{11}).  Thus, our present interpretation of $S^{BH}$ 
agrees with \cite{FF:97}.

4) Density matrix (\ref{11}) differs from the canonical density
matrix by insertion of the operator $\cal Q$ on $\Sigma$.
This insertion changes only the distribution of the soft
modes. The divergence of the thermal entropy $S$ in induced gravity
can be related to the degeneracy of states with respect to soft modes
\cite{FF:97} because adding a soft mode does not change the
canonical energy of the state. The insertion of $\cal Q$ in
(\ref{11}) removes this degeneracy. At the same time
averages of operators located outside  
$\Sigma$ do not depend on this insertion. 

{\bf Acknowledgements}:\ \ 
This work is supported in part by the RFBR grant
N 99-02-18146.

\end{document}